\documentclass{rmf-d}
\usepackage{nopageno,rmfbib,multicol,times,epsf,amsmath,amssymb,cite}
\usepackage[latin1]{inputenc}
\usepackage[]{caption2}
\usepackage{graphics}
\usepackage{mathtools}
\usepackage{cuted}
\usepackage{float}
\usepackage{mathrsfs}
\usepackage{color}

\definecolor{umich}{RGB}{0	, 0, 0}
\definecolor{umichR}{RGB}{0, 0, 0}

\newcommand{\e}{\textrm{e}}
\newcommand{\rF}{\textrm{F}}
\newcommand{\rC}{\textrm{C}}
\newcommand{\rV}{\textrm{V}}
\newcommand{\rS}{\textrm{S}}

\def\rmfcornisa{
\rmf\ 
\hfill December 2021
}

\def\rmfcintilla{{\it Rev.\ Mex.\ Fis.\/} {} }
\clearpage \rmfcaptionstyle 
\begin{document}
\title{   Graviton scattering amplitudes in first quantisation 
\vspace{-6pt}}
\author{ James P. Edwards     }
\address{  jpedwards@cantab.net   }

\maketitle

\begin{abstract}
\vspace{1em} We give a pedagogical review to alternative, first quantised approaches to calculating graviton scattering amplitudes, giving an introduction to string inspired approaches and presenting more recent work based on the worldline formalism of quantum field theory that is motivated by these  historic results. We describe how these first quantised techniques can greatly simplify the determination of such amplitudes, in particular reducing the number of Feynman-like diagrams that enter the computation and leading to compact results.  \vspace{1em}
\end{abstract}

\keys{  Quantum Field Theory, Gravitons, Scattering Amplitudes, Worldline Formalism  \vspace{-4pt}}

\begin{multicols}{2}

\section{Introduction}

The standard approach to perturbative quantum field theory (QFT), based on the Feynman diagram expansion that organises contributions according to the number of (virtual) loops, has been hugely successful, providing high-precision predictions for scattering processes involving fundamental particles. However, efforts to push calculations to higher loop orders are usually complicated by the factorial growth in the number of diagrams and the increasing difficulty in carrying out the resulting multi-dimensional integrals over virtual momenta. Despite this the resulting amplitudes often turn out to be much simpler than intermediate calculations would seem to anticipate and can involve orders of magnitude cancellations between diagrams to yield a relatively small final result. 

This is well illustrated by the famous example of the electron anomalous magnetic moment: Schwinger's seminal 1948 one-loop calculation \cite{Schwingerg-2} involves only one Feynman diagram, which was extended to the 7 diagram two-loop calculation just 9 years later \cite{Petermann, Sommerfeld}. It took until 1996 for the 72 three-loop diagrams to be calculated \cite{LaportaRemiddi}, while the four-loop result followed a majestic computation of 891 diagrams, published 21 years later \cite{Laporta} (some groups have analysed the five-loop diagrams numerically \cite{Kinoshita, Volkov}). Yet as is summarised in Table \ref{tabg-2}, the coefficients multiplying powers of the natural expansion parameter ($\frac{\alpha}{\pi}$, where $\alpha$, the fine structure constant, is defined in terms of the electric charge, $e$,  as $\alpha \equiv \frac{e^{2}}{4\pi}$) end up being close to small half-integers thanks to spectacular cancellations between Feynman diagrams. 
\begin{center}
\begin{table*}
\centering
\begin{tabular}{|c|c|c|c|}
\hline
\hspace{1em}\textbf{Order}\hspace{1em} & \hspace{1em}\textbf{Complexity}\hspace{1em}  & \hspace{1em}\textbf{Result (added to} $\mathbf{\frac{g-2}{2}}$\textbf{)} \hspace{1.5em}& \hspace{1em}\textbf{Timeline}\hspace{1em} \\ \hline
1-loop & 1 Diagram & $\frac{1}{2}\frac{\alpha}{\pi}$ & 1948 --- Schwinger \cite{Schwingerg-2}\\ \hline
2-loop & 7 Diagrams &$-0.328\ldots\big(\frac{\alpha}{\pi}\big)^2$& 1957 --- Petermann \cite{Petermann} / Sommerfeld \cite{Sommerfeld} \\ \hline
3-loop & 72 Diagrams &$+1.181\ldots\big(\frac{\alpha}{\pi}\big)^3$ & 1996 --- Laporta, Remiddi \cite{LaportaRemiddi} \\ \hline
4-loop & 891 Diagrams &$-1.912\ldots\big(\frac{\alpha}{\pi}\big)^4$ & 2017 --- Laporta \cite{Laporta} \\ \hline
5-loop & 12672 Diagrams & $--$ & $--$ \\ \hline
$\cdots$ & $\cdots$ & $\cdots$ & $\cdots$
\end{tabular}
\caption{Contributions to the electron $g-2$ at various loop orders (QED).}
\label{tabg-2}
\end{table*}
\end{center}
\vspace{-1.75em}

There are also cancellations of spurious UV divergences between diagrams influenced by gauge symmetry, which one might reasonably also connect to the subtle cancellations that leave behind such small finite parts mentioned above. Indeed, this has motivated Cvitanovi\'{c} to propose grouping diagrams into so-called ``gauge sets,'' whereby sets of gauge invariant diagrams are both UV finite and give contributions that are close to being integer multiples of $\pm \frac{1}{2}$ multiplied by the appropriate power of the perturbative expansion parameter\footnote{This raises the possibility of a softer growth of the (quenched) QED coefficients and perhaps a finite radius of convergence for this series \cite{Pred}.}. The issues of factorial growth and cancellation of divergences is all the more complicated in the case of graviton amplitudes, as we shall outline below, which is a significant motivation for the approaches presented in this contribution.

Indeed, these considerations suggest that it may be advantageous to consider alternative calculational techniques that avoid the Feynman diagram machinery. If such methods were better able to manifest the gauge symmetry of the theory, one may hope that intermediate calculations could be cleaner and it may be easier to understand the origins of the finiteness and numerical value of the final result. Here we present two such approaches, both of which based on \textit{first quantised} representations of field theory, as a pedagogical review: a \textit{string inspired} technique developed by Bern, Dunbar and Shimada (BDS) \cite{BDS}, extending the so-called Master Formula obtained for QCD by Bern and Kosower \cite{BK1, BK2} from an infinite tension limit of string theory to the case of graviton amplitudes; and the \textit{worldline formalism}, pioneered by Strassler \cite{Strass1} following initial suggestions by Feynman \cite{Feyn1, Feyn2}.

The outline of this contribution is as follows: in section \ref{secGrav} we discuss the difficulties in calculating graviton amplitudes in the standard approach and compare to the analogous scattering of photon and graviton states in open and closed string theory. We follow by outlining the worldline approach to photon scattering in QED in section \ref{secWL}, which will provide the base from which to present two alternatives to determining graviton amplitudes within first quantisation in section \ref{secBDS}. We end with some summarising conclusions.
\vspace{-1em}
\section{Graviton amplitudes}
\label{secGrav}
Continuing the thread of the introduction, the rapid growth in number and complexity of Feynman diagrams is even more apparent in the case of graviton amplitudes. This can be seen -- at least superficially -- by expanding the Einstein-Hilbert action, which to fix our conventions will be taken as
\begin{equation}
	S_{\textrm{EH}} = \frac{2}{\kappa^{2}} \int d^{D}x\, \sqrt{-g}R\,,
	\label{eqEH}
\end{equation}
where $R$ is the Ricci scalar, $\kappa^{2} = 32\pi G_{N}$ is the coupling constant derived from Newton's constant, $G_{N}$ and $g \equiv \det(g_{\mu\nu})$, about flat space, for which we set $g_{\mu\nu}(x) \rightarrow \eta_{\mu\nu} + \kappa h_{\mu\nu}(x)$. The complete diffeomorphism symmetry of the full action appears as a residual symmetry for the metric perturbation $h_{\mu\nu}$. If we work in de Donder gauge ($\partial^{\alpha}h_{\alpha\mu}- \frac{1}{2}\partial_{\mu}h = 0$ with $h \equiv \eta^{\mu\nu}h_{\mu\nu} = \tr(h)$) then the expansion of the action takes the following form:
\end{multicols}
\begin{align}
	S_{\textrm{EH}} =  \int d^{D}x\, \Big[\partial_{\mu}h_{\rho\sigma}\partial^{\mu}h^{\rho\sigma} - \frac{1}{2}\partial_{\mu}h\partial^{\mu}h + &\kappa\big(h_{\rho\sigma}\partial_{\mu}h^{\rho\sigma}\partial^{\mu}h - h_{\rho\sigma}\partial^{\rho}h_{\mu\nu}\partial^{\sigma}h^{\mu\nu}- 2h_{\rho\sigma}\partial_{\mu}h^{\rho}{}_{\nu}\partial^{\mu}h^{\nu\sigma} \nonumber \\
	+\frac{1}{2}& h \partial_{\mu} h_{\rho\sigma}\partial^{\mu}h^{\rho\sigma} + 2h_{\rho\sigma}\partial^{\mu}h^{\nu\rho}\partial^{\sigma}h_{\mu\nu} - \frac{1}{4}h \partial_{\mu}h \partial^{\mu}h\big) + \ldots\Big]\,,
	\label{eqEHExp}
\end{align}
\hspace{0.55\textwidth}\rule{0.4\textwidth}{0.15pt}
\begin{multicols}{2}
\noindent where the additional terms indicated by the ellipsis involve progressively higher orders in $h_{\mu\nu}$ (and the coupling, $\kappa$). 

Now the first two terms, quadratic in the metric perturbation, imply a graviton propagator (inverse to the $4$-index symmetric kinetic operator) as usual. It is a straightforward calculation to derive this propagator in momentum space, 
\begin{equation}
	P_{\mu\nu, \alpha\beta}(k) = \frac{1}{2}\frac{i}{k^{2} + i\epsilon}\Big[ \eta_{\mu\alpha}\eta_{\nu\beta} + \eta_{\mu\beta}\eta_{\nu\alpha} - \frac{2}{D-2}\eta_{\mu\nu}\eta_{\alpha\beta}\Big]\,.
	\label{eqProp}
\end{equation}
For later comparison to string theory, we note here that the final piece of the expression in brackets (trace term) makes the organisation of the perturbative expansion rather different from the way in which graviton amplitudes turn out on the string worldsheet where this piece is missing. This point, and a means of removing this part of the propagator are discussed in \cite{BDS}.

We deduce the (tree-level) multi-graviton vertices from the remaining terms in the expansion. It becomes clear that in contrast to QED ($3$-point vertex) or QCD ($3$- and $4$-point vertices), graviton amplitudes involve Feynman diagrams with an infinite number of vertices (that couple an arbitrary number of gravitons), whose tensor structures become progressively more complex -- indeed, even the simplest, three point vertex contains around $100$ terms in momentum space. Some examples of the multi-graviton vertices are illustrated in figure 1.
It is easy to see, then, that the Feynman diagram expansion for graviton amplitudes will be combinatorically far more complicated than in gauge theories. To give just a few examples, generic four graviton (e.g. $gg \rightarrow gg$) diagrams (see figure 2) will involve at least $\mathcal{O}(10^{20})$ terms already at $3$-loop order, rising to $\mathcal{O}(10^{26})$ by $4$-loop and to $\mathcal{O}(10^{31})$ terms at $5$-loop order, making such calculations essentially impossible using standard techniques.
\begin{figure}[H]
	\centering
	\includegraphics[width=0.525\textwidth]{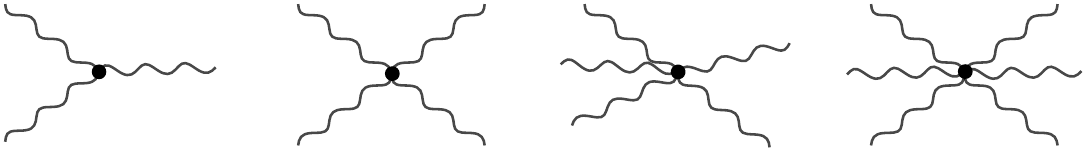}
	\caption{Examples of $3$-, $4$-, $5$- and $6$-point graviton vertices generated by expanding the Einstein-Hilbert action, (\ref{eqEH}), in the weak field limit -- see (\ref{eqEHExp}) for the terms providing the $3$-point contribution.}
	\label{figVertices}
\end{figure}
\vspace{-1.2em}
On the other hand, in recent years various alternative techniques for studying graviton amplitudes have been developed that strongly suggest they are simpler than might appear in the Lagrangian formulation. Double copy relations relating gravity to ``the square'' of gauge theory \cite{KLT, BernKLT}, originally uncovered in the context of (first quantised) string theory, combined with modern recursion relations, unitarity methods and related techniques \cite{BCF, BCFW, BDDK, CHY, BCJ} show that physical, \textit{on-shell} graviton amplitudes can be constructed from appropriate kinematic and colour factors derived from diagrams involving only \textit{3-point} vertices. In this contribution we shall return to older relations, again inspired by string theory, which arrive at the same conclusion, making both the enumeration of graviton amplitude diagrams and their eventual evaluation feasible.
\vspace{-0.95em}\begin{figure}[H]
	\centering
	\includegraphics[width=0.45\textwidth]{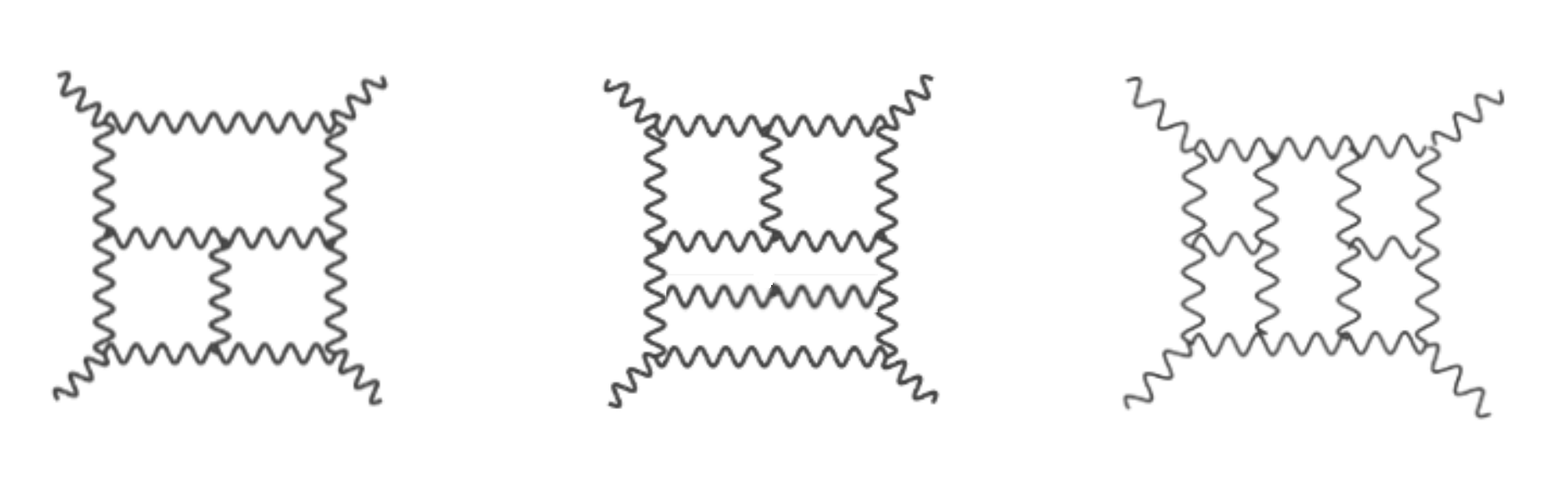}
	\caption{Examples of $3$-, $4$-, and $5$- loop diagrams for a $2\rightarrow 2$ graviton scattering process involving only $3$-graviton vertices.}
	\label{figDiagrams}
\end{figure}
\subsection{String theory amplitudes}
To understand the benefits of a string based approach, we recall that it is well-known that the infinite tension limit of string theory amplitudes is related to scattering amplitudes in corresponding field theories (below we shall give a precise example for photon scattering in the context of the worldline formalism of QFT) -- see, amongst others, \cite{Scherk, Yoneya, Schwarz}. Moreover, the reorganisation of field theory amplitudes within the string theory means that the contributions from multiple field theory Feynman diagrams can be obtained from a single string diagram, as illustrated in figure 3 for a generic theory.
\begin{figure}[H]
	\centering
	\vspace{-0.5em}
	\includegraphics[width=0.2\textwidth]{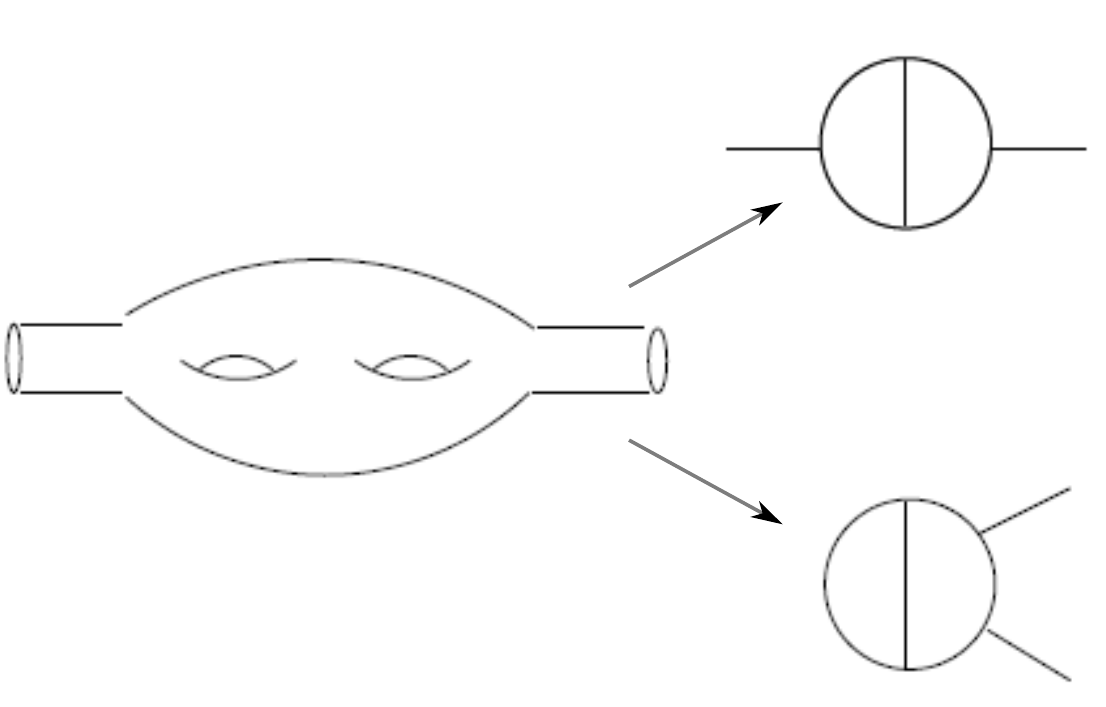}
	\caption{Schematic illustration of two-loop Feynman diagrams for $\phi^{3}$ theory produced in the infinite tension limit of a genus two string theory process -- adapted from \cite{ChrisRev}.}
	\label{figString}
\end{figure}
Before presenting the worldline description of photon scattering amplitudes inspired by this correspondence we revise here how scattering amplitudes between string states are calculated. Using Polyakov's representation of string theory (bosonic strings for simplicity) we write the amplitude as a path integral over worldsheet embeddings, $X\!\!: \Sigma \rightarrow R^{D}$, and geometries described by metrics $h$. Defining the worldsheet action by ($\alpha'$ is the inverse string tension)
\begin{equation}
	S[X, h]:= \frac{1}{4\pi \alpha'}\int_{\Sigma}d^{2}\sigma\, \sqrt{h}h^{\alpha\beta}\partial_{\alpha}X \cdot \partial_{\beta}X\,,
	\label{eqSPoly}
\end{equation}
an external string state is represented by a vertex operator $V(k, \varepsilon)$ under the path integral according to ($d^{2}\sigma \equiv d\tau d\sigma$)
\begin{equation}
\hspace{-1.75em}	\big \langle \prod_{i} V_{i}(k_{i}, \varepsilon_{i}) \big \rangle \sim \int \mathscr{D}h(\tau, \sigma)\int \mathscr{D}X(\tau, \sigma)\prod_{i} V_{i}(k_{i}, \varepsilon_{i})\, \e^{-S[X, h]}\,.
	\label{eqStringPI}
\end{equation}
The states in the amplitude should be from the string spectrum. For open strings, these could be the spin-zero scalar, $\phi$, (tachyon), or spin-one photon, $\gamma$, whose vertex operators involve integrals along the worldsheet boundary ($\sigma^{\pm}$, $\textrm{const.}$),
\begin{align}
	\hspace{-0.75em}V^{\phi}[k] &:= \int_{\partial\Sigma}d\tau \, \e^{i k \cdot X(\tau, \sigma^{\pm})}\,, \quad k^{2} = -\frac{1}{\alpha'}\,,\nonumber \\
	\hspace{-0.75em}V^{\gamma}[k, \varepsilon] &:= \int_{\partial \Sigma} d\tau \, \varepsilon \cdot \dot{X}(\tau, \sigma^{\pm})\, \e^{i k \cdot X(\tau, \sigma^{\pm})}\,, \quad k^{2} = 0 = k \cdot \varepsilon\,.
		\label{eqVerticesOpen}
\end{align}
The graviton (spin-two) is part of the closed string spectrum and its vertex operator allows this state to be inserted over the whole worldsheet (we have defined the combinations $\partial \equiv \partial_{\tau} + i\partial_{\sigma}$ and $\bar{\partial} \equiv \partial_{\tau} - i\partial_{\sigma}$)
\begin{align}
	V^{g}[k, \varepsilon]:= &\int_{\Sigma} d^{2}\sigma \, \partial X(\tau, \sigma) \cdot \varepsilon \cdot \bar{\partial} X(\tau, \sigma)\, \e^{i k \cdot X(\tau, \sigma)}\,, \nonumber \\
	 &\quad k^{2} = 0 = k \cdot \varepsilon = 0 = \varepsilon \cdot k\,.
	 \label{eqVertexg}
\end{align}
In the preceding vertex operators the mass-shell and transversality conditions follow from the requirement of evading the Weyl anomaly (in the critical dimension) -- see \cite{Mansfield}.

The path integral also sums over topologies of the worldsheet. With the above conditions satisfied, on a given Riemann surface the reparameterisation and conformal symmetries of the Polyakov theory, (\ref{eqSPoly}), allow the metric to be gauge fixed to be conformally flat  and, assuming the critical dimension, the path integral over metrics, $\int \mathscr{D}h(\tau, \sigma)$, eventually reduces to a Riemann integral over a finite space of conformal equivalence classes. On this gauge slice the matter path integral over $X(\tau, \sigma)$ is Gaussian and so (\ref{eqStringPI}) can be computed using Wick's theorem. Here we restrict attention to open strings -- on the annulus the fundamental contraction is the simple function inverting the Laplacian along its boundaries
\begin{align}
	\langle X^{\mu}(\tau_{1}) X^{\nu}(\tau_{2}) \rangle &\equiv \eta^{\mu\nu}G(\tau_{1} - \tau_{2} ; \tau)\\
	G(\tau_{1} - \tau_{2}; \tau) &=-\Big[ \log\big| 2 \sinh(\tau_{1} - \tau_{2}) \big| - \frac{(\tau_{1} - \tau_{2})^{2}}{\tau}  \nonumber \\
	& -4\e^{-2\tau }\sinh^{2}\big(\tau_{1} - \tau_{2}\big) \Big] + \mathcal{O}(q^{2})\,,
	\label{eqGString}
\end{align}
\noindent where the modular parameter, $q = \e^{-2\tau}$, the square of the ratio of the annulus' radii is written in terms of the length of the boundary, $\tau$, to be integrated over its fundamental domain. 

Now, following \cite{BK1, BK2} the infinite tension limit, $\alpha' \rightarrow 0$, corresponds to $\tau \rightarrow \infty$, $|\tau_{i} - \tau_{j}| \rightarrow \infty$, so that the ratio of the radii tends to $1$ (c.f. figure 3). In this limit the Green function and its derivative give the leading contributions
\begin{align}
	G(\tau_{1} - \tau_{2} ; \tau) &\sim \textrm{const} - \Big[ |\tau_{1} - \tau_{2}| - \frac{(\tau_{1} - \tau_{2})^{2}}{\tau} \Big] + \ldots\,, \nonumber\\
	\dot{G}(\tau_{1} - \tau_{2} ; \tau) &\sim -\Big[ \sigma(\tau_{1} - \tau_{2}) - 2\frac{\tau_{1} - \tau_{2}}{\tau} \Big] + \ldots\,.
	\label{eqGLimits}
\end{align}
Focussing on one-loop $N$-photon (easily extended to $N$-gluon \cite{BK1, BK2}) scattering, then, one can derive the Bern-Kosower (BK) rules that give a prescription for constructing an integral representation of the amplitude based on a Kinematic Factor, $\mathcal{K}_{N}$, derived from $\big\langle \prod_{i = 1}^{N}V^{\gamma}[k_{i}, \varepsilon_{i}]\big\rangle$ as in (\ref{eqStringPI})\footnote{One must also subtract contributions divergent as $q \rightarrow 0$ produced by tachyonic scalars running in the loop.},
\begin{equation}
	\hspace{-1.5em}\mathcal{K}_{N} \sim \int \prod_{i = 1}^{N}du_{i}\prod_{i<j} \exp\Big[ G_{ij}k_{i}\cdot k_{j}  + i\dot{G}_{ij}(k_{i} \cdot \varepsilon_{j} - k_{j}\cdot \varepsilon_{i}) + \ddot{G}_{ij}\varepsilon_{i} \cdot \varepsilon_{j} \Big]\,,
	\label{eqKDef}
\end{equation}
which plays the role of a kind of generating function for the one-loop amplitudes of figure 4; in fact the accompanying ``replacement rules'' (see below) allow this same Kinematic Factor to be reused to generate scattering amplitudes for various theories (i.e. different particles running in the loop).
\begin{figure}[H]
	\centering
	\includegraphics[width=0.425\textwidth]{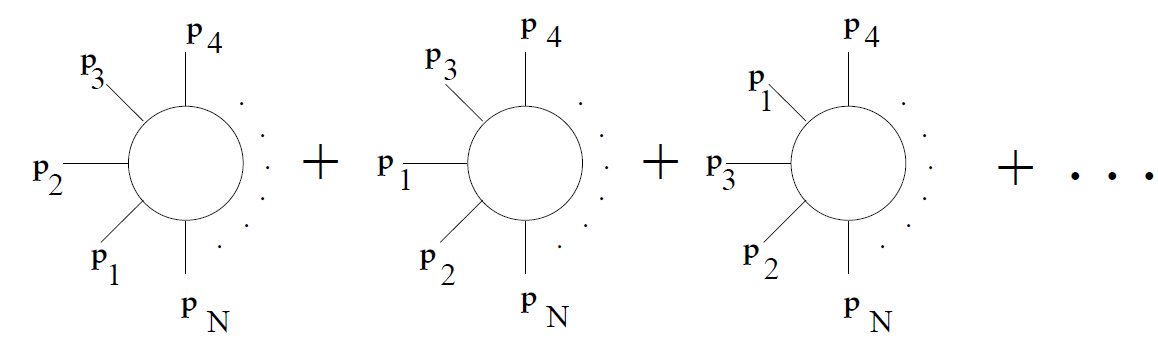}
	\caption{One-loop $N$-photon scattering amplitudes of the type produced by $\mathcal{K}_{N}$ -- so far agnostic regarding the particle in the loop.}
	\label{figPhotons}
\end{figure}
Even for photon / gluon amplitudes the Bern-Kosower rules have important advantages over perturbation theory. Combining multiple Feynman diagrams into one (figure 4) leads to a better organisation of gauge invariance, helped further by the fact that loop momentum integrals are already done leaving fewer kinematic invariants in intermediate calculations. It is also a universal basis for applying the replacement rules for different field theories that allow it to combine nicely with internal or space-time symmetries.
We shall explain the rules for manipulating the Kinematic Factor below, where we shall use it to generate graviton amplitudes based only on cubic vertices, but first we digress to explain how the Kinematic Factor can be derived purely within field theory.

\section{Worldline formalism}
\label{secWL}
The idea of a first quantised representation of field theory processes goes back to Feynman \cite{Feyn1, Feyn2} and development of what is now called the worldline approach in \cite{Strass1} was motivated by the BK results discussed above. Here we briefly describe the worldline formalism for one-loop $N$-photon amplitudes in scalar QED -- for reviews see \cite{ChrisRev, UsRep, 103}.

The effective action of scalar QED is defined by integrating out the matter degrees of freedom of the complex Klein-Gordon field, $\Phi$, that is minimally coupled to the Maxwell potential, $A_{\mu}$, according to (Euclidean space)
\begin{align}
\hspace{-1em}	\e^{\Gamma[A]} &\equiv \int \mathscr{D}\bar{\Phi}(x)\mathscr{D}\Phi(x) \, \e^{- \int d^{D}x \, \bar{\Phi}(x) (-D^{2} + m^{2}) \Phi(x)} \\
\hspace{-1em}	&= \textrm{Det}^{-1}\big(-D^{2} + m^{2}\big)\,,
	\label{eqGammaDef}
\end{align}
where $D_{\mu} \equiv \partial_{\mu} + i e A_{\mu}$ is the covariant derivative. Using the (functional) identity $\log \textrm{Det} (\hat{\mathcal{O}}) = \textrm{Tr}\log (\hat{\mathcal{O}})$ for operators $\hat{\mathcal{O}}$, with the Schwinger proper time trick to exponentiate the operator, we then evaluate the trace in position space:
\begin{align}
	\Gamma[A] &= -\textrm{Tr} \log\big(-D^{2} + m^{2}\big) \\
	&= \int_{0}^{\infty} \frac{dT}{T} \int d^{D}x \, \big\langle x \big| \e^{-T(-D^{2} + m^{2})} \big| x\big\rangle\,.
	\label{eqGammaInt}
\end{align}
The transition amplitude in the last line admits a natural path integral representation, over an auxiliary relativistic point particle, $x(\tau)$, traversing closed loops in proper time $T$, so 
\begin{align}
	\Gamma[A]  = \int_{0}^{\infty} \frac{dT}{T} e^{-m^{2}T} \oint_{PBC} \hspace{-0.5em}\mathscr{D}x(\tau)\, \e^{-S[x]}\,,
	\label{GammaPI}
\end{align}
where the worldline action, inherited from the evolution-like operator $\e^{-T(-D^{2})}$ in (\ref{eqGammaInt}), is given by
\begin{equation}
	S[x] = \int_{0}^{T}d\tau\, \Big[ \frac{\dot{x}^{2}}{4} + eA(x(\tau))\cdot \dot{x}(\tau) \Big]\,.
	\label{eqSWL}
\end{equation}
We can interpret (\ref{GammaPI}) as producing quantum corrections to the dynamics of the gauge field, generating all one-loop diagrams involving an arbitrary number of couplings to $A_{\mu}$. 
\vspace{-0.35em}
\subsection{Photon amplitudes}
\vspace{-0.25em}The one-loop $N$-photon amplitudes are extracted from the effective action by specialising the background field to a sum of plane waves representing external states of fixed polarisation, $\varepsilon_{i}$, and momentum, $k_{i}$, so that 
\begin{equation}
	A_{\mu}(x) = \sum_{i = 1}^{N}\varepsilon_{i\mu}\e^{i k_{i} \cdot x}\,,
	\label{eqAPhotons}
\end{equation}
and then selecting from $\Gamma[A]$ the part multi-linear in the polarisations. This provides a path integral representation of the amplitudes in analogy to the string theory case, (\ref{eqStringPI}),
\begin{equation}
	\hspace{-2em} \Gamma_{N}[\{k_{i}, \varepsilon_{i}\}] = (-ie)^{N}\int_{0}^{\infty}\frac{dT}{T}\e^{-m^{2}T}  \oint_{PBC} \hspace{-0.5em}\mathscr{D}x(\tau)\,\e^{- \int_{0}^{T}\frac{\dot{x}^{2}}{4}} \prod_{i = 1}^{N}V^{\gamma}[k_{i}, \varepsilon_{i}]\,,
\end{equation}
where the vertex operator has the same form as in string theory, (\ref{eqVerticesOpen}), but now integrating along the particle's trajectory:
\begin{equation}
	V^{\gamma}[k, \varepsilon] = \int_{0}^{T}d\tau\, \varepsilon\cdot \dot{x}(\tau) \e^{i k \cdot x(\tau)}\,.
	\label{eqVertexWL}
\end{equation}
Note, however, that we have \textit{not} been forced to impose on-shell or transversality conditions. At this stage the path integral is Gaussian, so we can evaluate it using Wick's theorem after separating off the constant zero mode. Expanding about the loop centre of mass, $x^{\mu}(\tau) \rightarrow x_{0}^{\mu} + q^{\mu}(\tau)$ replaces $\oint_{\textrm{PBC}} \mathscr{D}x(\tau) \longrightarrow \int d^{D}x_{0} \int_{\textrm{SI}}\mathscr{D}q(\tau)$ with ``String-Inspired'' boundary conditions on the deviation, $q^{\mu}(0) = 0 = q^{\mu}(T)$ and $\int_{0}^{T}d\tau\, q^{\mu}(\tau) = 0$, and in this space orthogonal to the zero mode the Green function for the kinetic term is
\begin{align}
	\hspace{-1em}\big\langle q^{\mu}(\tau_{i})q^{\nu}(\tau_{j})\big\rangle_{\perp} &= -G_{Bij}\eta^{\mu\nu} \,,\\ 
	\hspace{-1em}\quad G_{Bij} \equiv G_{B}(\tau_{i}, \tau_{j}) &= |\tau_{i} - \tau_{j}| - \frac{(\tau_{i} - \tau_{j})^{2}}{T}\,.
	\label{eqGreenWL}
\end{align}
Note that this Green function coincides with the leading order behaviour of the string theory Green function in (\ref{eqGLimits}) up to an irrelevant constant. For a formal determination of the path integral we borrow yet another string theory trick, exponentiating the prefactor of the vertex operator as $V^{\gamma}[k, \varepsilon] = \int_{0}^{T}d\tau\,  \e^{i k \cdot x(\tau) + \varepsilon\cdot \dot{x}(\tau)}\big|_{\varepsilon}$, retaining only the linear part in $\varepsilon$. Then, completing the square in the exponent of the path integral we recover the Bern-Kosower Master Formula, still valid even off-shell, containing the Kinematic Factor, $\mathcal{K}_{N}$
\begin{align}
	\hspace{-1.5em}\Gamma_{N}[\{k_{i},\varepsilon_{i}\}] &= (-ie)^{N}(2\pi)^{D}\delta^{D}\big(\sum_{i}k_{i}\big)\! \int_{0}^{\infty}\frac{dT}{T}(4\pi T)^{-\frac{D}{2}}\e^{-m^{2}T}\nonumber\\
	&\hspace{-3em} \prod_{i=1}^{N}\int_{0}^{T}d\tau_{i} \,\e^{\frac{1}{2}\sum_{i, j=1}^{N} G_{Bij}k_{i}\cdot k_{j} -2i \dot{G}_{Bij}\varepsilon_{i}\cdot k_{j} + \ddot{G}_{Bij}\varepsilon_{i}\cdot \varepsilon_{j}}\Big|_{\varepsilon_{1}\ldots \varepsilon_{N}},
	\label{eqBKMF}
\end{align}
where the momentum conserving $\delta$-function arose from integrating over $x_{0}$. The notation at the end of the second line indicates that one should expand to multi-linear order in the $\varepsilon_{i}$. After this, the eventual integral over proper time, $T$, produces the familiar Feynman parameter denominator ${\big[m^{2} -\sum_{i < j = 1}^{N} k_{i}\cdot k_{j} G_{Bij}\big]^{\frac{D}{2} - N}}$, but in a way that unifies the different orderings of insertions of the external photons around the loop -- see figure 4, or \cite{UsSigma} for progress exploiting this property under the parameter integrals. Having thus shown how to arrive at the Kinematic Factor, we shall now describe its generalisation to graviton amplitudes and attempts to extend the procedure to off-shell processes.

\section{Bern-Dunbar-Shimada Rules}
\label{secBDS}
The extension of the Bern-Kosower rules to gravity was systematically studied in \cite{BDS}, building upon \cite{BGK,GSB} and was subsequently applied by Dunbar and Norridge to determine $4$-graviton amplitudes at one-loop order for all helicity assignments \cite{DunNor}. Here we recapitulate their construction before mentioning some efforts towards extending the technique using the worldline formalism.

The major difference with respect to photon or gluon amplitudes is the graviton vertex operator, (\ref{eqVertexg}), inserted on closed string worldsheets where there are two ``sectors'' that contribute to the amplitude, corresponding to left- and right-moving string modes. This also implies that the worldsheet Green function (analogous to (\ref{eqGString})) becomes a genuine function of two variables, $\sigma^{\pm} := \tau \pm i\sigma$, for these sectors. Then starting from $G(\sigma^{+}, \sigma^{-})$ we follow the notation of \cite{BDS, DunNor}:
\begin{itemize}
	\item We use $\dot{G}$ and $\ddot{G}$ for $\sigma^{+}$ derivatives of $G$.
	\item We use $\dot{\bar{G}}$ and $\ddot{\bar{G}}$ for $\sigma^{-}$ derivatives of $G$.
	\item We denote by $H$ the derivative of $G$ with respect to one left- and one right-moving variable.
\end{itemize}
Finally, we also decompose the on-shell graviton polarisation tensor, $\varepsilon$, into the two sectors by setting $\varepsilon_{\mu\nu} \longrightarrow \varepsilon_{\mu}\bar{\varepsilon}_{\nu}$ and later reconstruct it by identifying $\varepsilon_{\mu}\bar{\varepsilon}_{\nu} \equiv \varepsilon_{\mu\nu}$ at the end.

\subsection{$N$-graviton rules}
The one-loop $N$-graviton amplitudes are then generated from some ``primordial Feynman diagrams'' involving only the cubic vertices produced by the string splitting process according to the following (simplified) steps developed in \cite{BDS}:

\medskip
\noindent \underline{\textbf{Step 1:}}\\
Draw all one-loop diagrams having $\Phi^{3}$ topology with $N$ external legs with appropriate labels, such as those in figure 5.
\begin{figure}[H]
	\centering
	\includegraphics[width=0.42\textwidth]{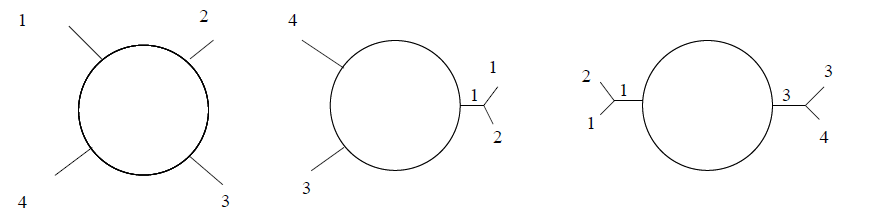}
	\caption{Primordial $\Phi^{3}$ diagrams for a $4$-graviton process \cite{BDS}.}
	\label{figPhi3}
\end{figure}
\noindent All permutations of external legs should be included and labelled as in conventional perturbation theory (in contrast to gluon amplitudes there is no need to worry about colour ordering). To internal legs attached to ``external trees,'' assign a label equal to the smallest label of the external legs it opens up to. However, we \textit{do} ignore ``tadpole'' diagrams or diagrams involving loops on \textit{external} legs such as in figure 6.
\begin{figure}[H]
	\centering
	\includegraphics[width=0.275\textwidth]{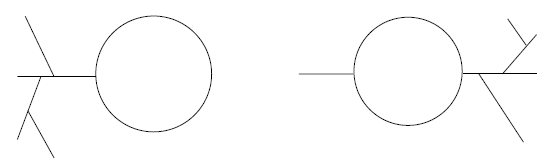}
	\caption{Tadpoles and isolated loops on external legs are ignored (they are renormalised or vanish in dimensional regularisation).}
	\label{figTad}
\end{figure}
\noindent \underline{\textbf{Step 2:}}\\
We calculate the contribution from each diagram by a reduction process. To each diagram we associate an integral (in dimensional regularisation we take $D = 4-2\epsilon$)
\end{multicols}
\rule{0.4\textwidth}{0.15pt}
\begin{equation}
		\mathscr{D} = i\frac{(-\kappa)^{N}}{(4\pi)^{2-\epsilon}}\Gamma\big[\ell - 2 + \epsilon\big] \int_{0}^{1}du_{\ell  - 1} \int_{0}^{u_{\ell - 1}}\!dx_{\ell - 2}\cdots \int_{0}^{u_{2}}du_{1} 
		 \frac{\mathcal{K}_{\textrm{red}}}{\big[ \sum_{i<j} K_{i}\cdot K_{j} G_{ij}  \big]^{\ell - 2 + \epsilon}}\,,
	\label{eqDDef}
\end{equation}
\hspace{0.55\textwidth}\rule{0.4\textwidth}{0.15pt}
\begin{multicols}{2}
\noindent where $l$ is the number of lines attached to the (massless) loop. Here the ordering of the parameter integrals over the $u_{i}$ should match the ordering of these lines about the loop and we have introduced the momenta $K_{i}$ entering the loop at point $i$, being just the sum of the external momenta entering the trees that join to the loop there. Finally, the graviton Reduced Kinematic Factor, $\mathcal{K}_{\textrm{red}}$, that survives the field theory limit of string theory is to be constructed from the generalised Kinematic Factor formed by multiplying contributions from the left- and right-moving sectors:
\begin{align}
\hspace{-1em}	\mathcal{K}_{N} = \int \prod_{i=1}^{N}du_{i}d\bar{u}_{i}\prod_{i<j}^{N}\e^{k_{i}\cdot k_{j} G_{ij}} 	\,&\e^{(k_{i}\cdot \varepsilon_{j} - k_{j}\cdot \varepsilon_{i})\dot{G}_{ij} - \varepsilon_{i}\cdot \varepsilon_{j} \ddot{G}_{ij}} \nonumber \\
\hspace{-1em}\vspace{-2em}	&\e^{(k_{i}\cdot \bar{\varepsilon}_{j} - k_{j}\cdot \bar{\varepsilon}_{i})\dot{\bar{G}}_{ij} - \bar{\varepsilon}_{i}\cdot \bar{\varepsilon}_{j} \ddot{\bar{G}}_{ij}} \nonumber \\
\hspace{-1em}	&\e^{-(\varepsilon_{i}\cdot \bar{\varepsilon}_{j} + \varepsilon_{j}\cdot \bar{\varepsilon}_{i})H_{ij} }\Big|_{\varepsilon_{1}\bar{\varepsilon}_{1}\cdots \varepsilon_{N}\bar{\varepsilon}_{N}}
	\label{eqKGravity}
\end{align}

\noindent which is somewhat reminiscent of the double copy relations discussed above (see \cite{Ahmadiniaz:2021ayd} for an in depth study of this relation). This Factor is to be expanded to multi-linear order in each of the $\varepsilon_{i}$ and $\bar{\varepsilon}_{i}$ and is reduced to determine the Reduced Kinematic Factor in the following step:

\medskip
\noindent \underline{\textbf{Step 3:}} Integration By Parts \\
After expanding $\mathcal{K}_{N}$ to multi-linear order, we integrate by parts to remove all $\ddot{G}_{ij}$ and $\ddot{\bar{G}}_{ij}$ -- this is what makes it possible to reduce the calculation of the amplitude to diagrams with purely cubic vertices. In this process the functions $G_{ij}$ and its second derivatives are taken to be symmetric in their indices whilst first derivatives are anti-symmetric. The ``crossed derivatives'' are handled according to the relations
\begin{align}
	\hspace{-0.5em}	\frac{\partial}{\partial u_{k}}\dot{\bar{G}}_{ij}&=( \delta_{ki}- \delta_{kj})H_{ij}\,, \quad
		\frac{\partial}{\partial \bar{u}_{k}}\dot{G}_{ij}=( \delta_{ki}- \delta_{kj})H_{ij}\,, \nonumber \\
	\hspace{-0.5em}	\frac{\partial}{\partial u_{k}}\ddot{\bar{G}}_{ij}&=0 \,, \qquad \qquad  \qquad \quad
   	\!\!\frac{\partial}{\partial \bar{u}_{k}}\ddot{G}_{ij}= 0\,.
   	\label{eqCrossedDerivs}
	\end{align}
Achieving this, the leading exponential factor in (\ref{eqKGravity}) involving the $G_{ij}$ and the parameter integrals can be dropped (they are already included in $\mathscr{D}$), which leaves behind $\mathcal{K}_{\textrm{red}}$. We now transform this according to so-called \textit{replacement rules:}

\medskip
\noindent \underline{\textbf{Step 4a:}} Tree Replacement Rules\\
The particle loop may have external legs attached via trees, whose ``branches'' we now remove. Working from the \textit{outside} in, we pinch away the trees by the replacement
\begin{equation}
	(\dot{G}_{ij})(\dot{\bar{G}}_{ij}) \longrightarrow \frac{1}{2k_{i}\cdot k_{j}}\, \qquad (i < j)\,,
	\label{eqTreeReplace}
\end{equation}
replacing other powers of these derivatives to zero -- see figure 7. In the remaining expression, set $j \rightarrow i$ in any other factors of the $G_{jk}$ and iterate until only the loop remains; this isolates the on-shell poles in the $S$-matrix.
\vspace{-0.5em}\begin{figure}[H]
	\centering
	\def\svgwidth{0.2\textwidth}
	\hspace{-2em}\vspace{-0.5em}
\begingroup%
  \makeatletter%
  \providecommand\color[2][]{%
    \errmessage{(Inkscape) Color is used for the text in Inkscape, but the package 'color.sty' is not loaded}%
    \renewcommand\color[2][]{}%
  }%
  \providecommand\transparent[1]{%
    \errmessage{(Inkscape) Transparency is used (non-zero) for the text in Inkscape, but the package 'transparent.sty' is not loaded}%
    \renewcommand\transparent[1]{}%
  }%
  \providecommand\rotatebox[2]{#2}%
  \newcommand*\fsize{\dimexpr\f@size pt\relax}%
  \newcommand*\lineheight[1]{\fontsize{\fsize}{#1\fsize}\selectfont}%
  \ifx\svgwidth\undefined%
    \setlength{\unitlength}{27.08908441bp}%
    \ifx\svgscale\undefined%
      \relax%
    \else%
      \setlength{\unitlength}{\unitlength * \real{\svgscale}}%
    \fi%
  \else%
    \setlength{\unitlength}{\svgwidth}%
  \fi%
  \global\let\svgwidth\undefined%
  \global\let\svgscale\undefined%
  \makeatother%
  \begin{picture}(1,0.55199029)%
    \lineheight{1}%
    \setlength\tabcolsep{0pt}%
    \put(0,0){\includegraphics[width=\unitlength,page=1]{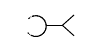}}%
    \put(0.22086141,0.26261446){\color[rgb]{0,0,0}\makebox(0,0)[lt]{\lineheight{1.25}\smash{\begin{tabular}[t]{l}LOOP\end{tabular}}}}%
    \put(0,0){\includegraphics[width=\unitlength,page=2]{Replacement.pdf}}%
    \put(0.57343131,0.30501016){\makebox(0,0)[lt]{\lineheight{1.25}\smash{\begin{tabular}[t]{l}$i$\end{tabular}}}}%
    \put(0.67119912,0.36470934){\makebox(0,0)[lt]{\lineheight{1.25}\smash{\begin{tabular}[t]{l}$i$\end{tabular}}}}%
    \put(0.67119912,0.18474798){\makebox(0,0)[lt]{\lineheight{1.25}\smash{\begin{tabular}[t]{l}$j$\end{tabular}}}}%
    \put(0.81079419,0.27062353){\makebox(0,0)[lt]{\lineheight{1.25}\smash{\begin{tabular}[t]{l}$({\color{umich}\dot{G}_{ij}})({\color{umichR}\dot{\bar{G}}_{ij}})\longrightarrow \frac{1}{2 k_{i}\cdot k_{j}}$\end{tabular}}}}%
    \put(0,0){\includegraphics[width=\unitlength,page=3]{Replacement.pdf}}%
  \end{picture}%
\endgroup%

	\caption{The Tree Replacement Rule pinches off a branch of a tree attached to the loop.}
	\label{figTreeReplace}
\end{figure}
\vspace{-0.25em}So far we have not been specific about the type of particle running in the loop,  fixed with the next replacement rules.

\medskip
\noindent \underline{\textbf{Step 4b:}} Loop Replacement Rules\\
To fix the field theory coupled to the gravitons we now transform the Reduced Kinematic Factor depending on which particle(s) will circulate in the loop. These rules are in fact independent implementations of the BK Replacement Rules for gauge theory amplitudes in the left- and right-moving sectors. 

For the simplest case of a scalar running in the loop, the replacements correspond to reintroducing the worldline Green function, (\ref{eqGreenWL}), after rescaling $\tau_{i} \rightarrow Tu_{i}$, as follows:
\begin{align}
		G_{ij} &\rightarrow G_{Bij} = |u_{i} - u_{j}| - (u_{i} - u_{j})^{2} = (u_{i} - u_{j})(1 - (u_{i} - u_{j})) \nonumber \\	
		\dot{G}_{ij} &\rightarrow -\frac{1}{2}\dot{G}_{Bij} = -\frac{1}{2}(\sigma(u_{i}-u_{j}) -2(u_{i}-u_{j})) \nonumber \\
		\dot{\bar{G}}_{ij} &\rightarrow -\frac{1}{2}\dot{G}_{Bij} = -\frac{1}{2}(\sigma(u_{i}-u_{j}) -2(u_{i}-u_{j}) \nonumber)\\
		H_{ij} &\rightarrow \frac{1}{2T}\,,
	\label{eqScalLoopReplace}
\end{align}
where we used the ordering of parameter integrals in (\ref{eqDDef}) and ignored a $\delta$-function in $H_{ij}$ that does not contribute on-shell. For a complex scalar running in the loop, $\mathcal{K}_{\textrm{red}}$ should be multiplied by $2$ for degrees of freedom. Now at this stage, the Reduced Kinematic Factor has been transformed to a genuine function of the external momenta and parameters $u_{i}$, so it can be substituted into $\mathscr{D}$ for the diagram in question to compute its contribution to the amplitude. 

We should note that the Loop Replacement Rules can be generalised to allow other particles in the loop. They are conveniently described with the notation \cite{DunNor} $\rF = \rS + \rC_{\rF}$ and $\rV = \rS + \rC_{\rV}$, where $\rS$ stands for the scalar Loop Replacement, (\ref{eqScalLoopReplace}), and $\rC_{\rF}$ and $\rC_{\rF}$ are \textit{Cycle Replacement Rules} that act on ``closed cycles'' of $\dot{G}$ and $\dot{\bar{G}}$ such as $\dot{G}_{ij}\dot{G}_{jk}\cdots \dot{G}_{si}$ to transform them into functional expressions\footnote{For $\rC_{\rV}$ the substitution is $G_{i_{1}i_{2}}G_{i_{2}i_{3}}\cdots G_{i_{n}i_{1}} \longrightarrow \frac{1}{2}\big(1 + \delta_{n,2}\big)$. For $\rC_{\rF}$ it is $G_{i_{1}i_{2}}G_{i_{2}i_{3}}\cdots G_{i_{n}i_{1}} \longrightarrow -\big(-\frac{1}{2}\big)^{n} \prod_{k = 1}^{n}\sigma(u_{i_{k}} - u_{i_{k+1}})$.}.
\begin{table}[H] \vspace{-0.5em}
\begin{tabular}{|c|c|}
\hline \hspace{1em}\textbf{Replacement Rule} \hspace{1em} & \hspace{1em}\textbf{Field Theory}\hspace{1em} \\ 
\hline [$\rS$, $\rS$] & Real Scalar \\
\hline 2[$\rS$, $\rS$] & Complex Scalar \\
\hline -2[$\rS$, $\rF$] & Weyl Fermion \\
\hline 2[$\rS$, $\rV$] & Vector Boson \\
\hline -4[$\rV$, $\rF$] & Gravitino $+$ Weyl Fermion \\
\hline 4[$\rV$, $\rV$] & Graviton $+$ Complex Scalar \\
\hline 4[$\rV$, $\rV$] - 2[$\rS$, $\rS$] & Graviton \\
\hline -4[$\rV$, $\rF$] + 2[$\rS$, $\rF$] & Gravitino \\
\hline
\end{tabular}
\caption{Loop replacement rules, based on transformations  $\rF = \rS + \rC_{\rF}$ and $\rV = \rS + \rC_{\rV}$, where $[\textrm{A}, \textrm{B}]$ indicates the replacements $\textrm{A}$ in the left-moving $\dot{G}$ sector and $\textrm{B}$ in the right-moving $\dot{\bar{G}}$ sector. }
\label{tabLoopReplace}
\end{table}

\vspace{-1.5em}
\subsection{Example application}
As this submission is a pedagogical introduction, we present here an application of the rules to $4$-graviton scattering, describing how the amplitude $\mathscr{A}(1,2,3,4)$ was calculated in \cite{BDS} for an illustrative helicity assignment. Actually, the rules combine nicely with spinor helicity techniques \cite{ElvangHuang} applied to the graviton polarisation tensor, where for fixed helicity we decompose $\varepsilon_{\mu\nu}^{\pm\pm} \rightarrow \varepsilon_{\mu}^{\pm} \bar{\varepsilon}_{\nu}^{\pm}$ --- then good choices of the references spinors for the $\varepsilon^{\pm}$ and $\bar{\varepsilon}^{\pm}$ can significantly simplify the exponent of $\mathcal{K}_{N}$ and lead to more compact expressions.

We limit ourselves to the Maximally Helicity Violating amplitude $\mathscr{A}(1^{-}, 2^{+}, 3^{+}, 4^{+})$, which in the standard formalism would involve 12 types of diagram totalling $54$ diagrams based on vertices with $\mathcal{O}(100)$ terms. With the string inspired approach and judicious choice of reference vectors that is reduced to just the $5$ diagrams with $\Phi^{3}$ topology, illustrated in figure 8, that will have the appropriate factors of $\dot{G}\dot{\bar{G}}$ to survive the Tree Replacement Rules (\textbf{Step 4a}). 
\begin{figure}[H]
	\centering
	\includegraphics[width=0.5\textwidth]{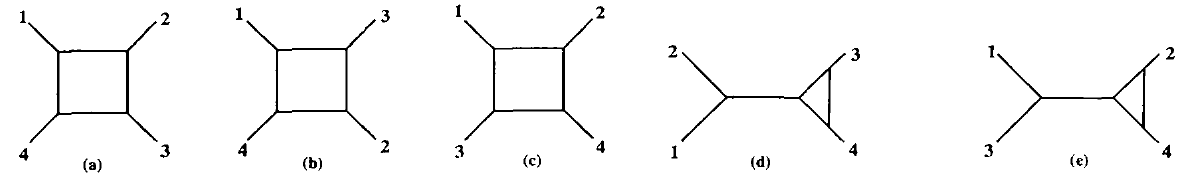}
	\caption{The appropriately labelled five diagrams that contribute to the $4$-graviton process $\mathscr{A}(1^{-}, 2^{+}, 3^{+}, 4^{+})$ \cite{DunNor}.}
	\label{figGraviton5}
\end{figure}
\vspace{-1em}The Kinematic Factor for these diagrams does not contain any second derivatives so we immediately get:
\begin{align}
		\hspace{-1em}\mathcal{K}_{\textrm{red}} = \mathscr{S}&(\dot{G}_{13}-\dot{G}_{12})(\dot{G}_{24}-\dot{G}_{23})(\dot{G}_{34}+\dot{G}_{23})(\dot{G}_{34}-\dot{G}_{24}) \nonumber \\
		\hspace{-1em}\times&(\dot{\bar{G}}_{13}-\dot{\bar{G}}_{12})(\dot{\bar{G}}_{24}-\dot{\bar{G}}_{23})(\dot{\bar{G}}_{34}+\dot{\bar{G}}_{23})(\dot{\bar{G}}_{34}-\dot{\bar{G}}_{24})\,,
	\label{eqKRed4}		
\end{align}
\vspace{-0.1em}where $\mathscr{S} = \left(\frac{s^{2}t}{4}\right)^{2} \left(\frac{[24]^2}{[12]\langle 23\rangle \langle 34 \rangle [41]}\right)^{2}$. We show how to apply the BDS rules to two of these diagrams to arrive at a compact, simple Lorentz invariant expression for this amplitude.

\medskip
\noindent\begin{tabular}{lc}\underline{\textbf{Diagram (a)}} & \raisebox{-2em}{\includegraphics[width=0.08\textwidth]{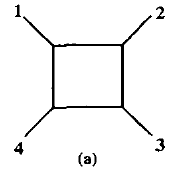}}\end{tabular}\\
Since there are no trees we can move directly to the Loop Replacement rules (\textbf{Step 4b}). For scalars, these lead to
\begin{equation}
	\mathcal{K}_{\textrm{red}}^{(a)} = 2\mathscr{S}u_{2}^{2}(1-u_{3})^{2}(u_{3} - u_{2})^{4}\,.
\end{equation}
For this diagram $\mathscr{D}$ is finite in $D = 4$ so in terms of the traditional Mandelstam variables we have
\begin{equation}
\hspace{-2.5em}	\mathscr{D}_{a} = \frac{2i\kappa^{4}}{(4\pi)^{2}} \mathscr{S} \int_{0}^{1}\!du_{3}\!\int_{0}^{u_{3}}\!du_{2}\!\int_{0}^{u_{2}}\!du_{1} \frac{u_{2}^{2}(1-u_{3})^{2}u_{32}^{4}}{\big[ s u_{1}u_{32} + tu_{21}(1 - u_{3}) \big]^{2}}
	\label{eqDa}
\end{equation}
where we used the shorthand $u_{ij} = u_{i} - u_{j}$. Calculating the integral and repeating the process for the similar diagrams (b) and (c) it is straightforward to verify the results
\begin{align}
	\hspace{-2.25em}\mathscr{D}_{a} = \frac{2i\kappa^{4}}{(4\pi)^{2}} \frac{\mathscr{S}}{840st}\,, \quad \mathscr{D}_{b} =\frac{2i\kappa^{4}}{(4\pi)^{2}} \frac{\mathscr{S}}{840ut}\,, \quad \mathscr{D}_{c} = \frac{2i\kappa^{4}}{(4\pi)^{2}} \frac{\mathscr{S}}{252su}\,.
	\label{eqDabc}\vspace{-1em}
\end{align}
\vspace{-0.25em}\noindent\begin{tabular}{lc}\underline{\textbf{Diagram (d)}} & \raisebox{-1.9em}{\includegraphics[width=0.115\textwidth]{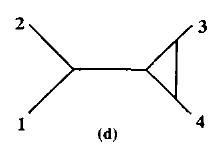}}\end{tabular}\\
This time there is a $1$-$2$ tree attached to the loop, so the Tree Replacement Rules (\textbf{Step 4a}) are invoked on the term in $\mathcal{K}_{\textrm{red}}$ involving $\dot{G}_{12}\dot{\bar{G}}_{12}$, transforming (\ref{eqKRed4}) by
\begin{align}
\mathcal{K}_{\textrm{red}}\rightarrow -\frac{\mathscr{S}}{s}&(\dot{G}_{24}-\dot{G}_{23})(\dot{G}_{34}+\dot{G}_{23})(\dot{G}_{34}-\dot{G}_{24})\nonumber \\
\times&(\dot{\bar{G}}_{24}-\dot{\bar{G}}_{23})(\dot{\bar{G}}_{34}+\dot{\bar{G}}_{23})(\dot{\bar{G}}_{34}-\dot{\bar{G}}_{24})\,.
	\label{eqKRedd}	
\end{align}
The Loop Replacement Rule (\textbf{Step 4b}) is now applied to turn this into an authentic function, which yields the finite integral
\begin{equation}
	\mathscr{D}_{d} = -\frac{2i\kappa^{4}}{(4\pi)^{2}}\frac{\mathscr{S}}{s} \int_{0}^{1}du_{3}\int_{0}^{u_{3}}du_{2} \frac{u_{2}^{2}(1-u_{3})^{2}(u_{3} - u_{2})^{2}}{s (u_{3} - u_{2})}\,.
	\label{eqDd}
\end{equation}
Evaluating this integral and repeating the process for diagram (e) provides the partial amplitudes
\begin{equation}
	\mathscr{D}_{d} = \frac{2i\kappa^{4}}{(4\pi)^{2}}\frac{\mathscr{S}}{360s^{2}} \,, \quad \qquad\mathscr{D}_{e} = \frac{2i\kappa^{4}}{(4\pi)^{2}}\frac{\mathscr{S}}{360u^{2}}\,.
	\label{eqDde}
\end{equation}
Summing up these results, then, the procedure has successfully determined the complete amplitude to be \cite{DunNor}
\begin{equation}
	\hspace{-1.25em}	\mathscr{A}(1^{-}, 2^{+}, 3^{+}, 4^{+}) = \frac{i\kappa^{4}}{(4\pi)^{2}}\frac{s^{2}t^{2}}{2880u^{2}}(u^{2} - st)\left(\frac{[24]^2}{[12]\langle 23\rangle \langle 34 \rangle [41]}\right)^{2}\,,
		\label{eqAFinal}
\end{equation} 
which would be far more difficult to get using standard techniques. Checks that (\ref{eqAFinal}) is consistent with appropriate crossing relations, symmetries, unitarity and other constraints \cite{DunNor} show this method to be a powerful alternative that bypasses prohibitively complicated field theory calculations.

\subsection{Worldline approach}
Despite its successes there are some drawbacks, most notably the requirement that the gravitons be on-shell, built into the string theory early on (in contrast to the photon / gluon case where it does not really matter in the infinite tension limit, as we saw in the worldline formalism in section \ref{secWL}) and the fact that we specialised to massless field theories. 

Worldline attempts to generalise the BDS construction to off-shell amplitudes with massive particles in the loop focus on the irreducible diagrams, since BDS can produce the reducible contributions. The worldline representation of one-loop $N$-graviton amplitudes for the scalar case is \cite{WLGrav1, WLGrav2}
\begin{equation}
	\hspace{-0.75em}\frac{1}{2}\left(\frac{-\kappa}{4}\right)^{N}\int_{0}^{\infty}\frac{dT}{T}(4\pi T)^{-\frac{D}{2}}\e^{-m^{2}T} \big\langle V^{g}[k_{1}, \varepsilon_{1}]\ldots V^{g}[k_{N}, \varepsilon_{N}] \big\rangle\,,
	\label{eqGravitonsWL}
\end{equation}
where the graviton vertex operator is, similarly to (\ref{eqVertexg}),
\begin{equation}
	V^{g}[k, \varepsilon] = \int_{0}^{T}d\tau\, \dot{x}\cdot \varepsilon\cdot \dot{x}\, \e^{ik\cdot x}\,.
	\label{eqVertexGravitonWL}
\end{equation}
Note that, unlike in string theory, there is no intrinsic separation of the vertex into left- and right-moving modes. Efforts to mimic this on the worldline and exploit integration by parts algorithms and other worldline techniques are hoped to permit an efficient extension of the BDS procedure that will be a viable alternative tool for studying graviton amplitudes. 
\vspace{-0.3em}
\section{Conclusion}
We have discussed the complexity of calculating amplitudes using standard perturbation theory to motivate alternative approaches that simplify their determination and then presented two string inspired techniques that, at least partially, realise this. For gauge theories such as QED or QCD, we already understand the string-based Bern-Kosower method via the worldline formalism, and we have described how this same formalism may be able to shed new light on the Bern-Dunbar-Shimada rules for graviton amplitudes. The advantages of producing Master Formulas that combine various Feynman diagrams and unify different field theories, distinguished only by appropriate Replacement Rules, are clear. Ongoing work on the worldline should extend these rules to massive, off-shell amplitudes in the same unifying framework. 
\end{multicols}
\vspace{-2.5em}
\medline
\vspace{-0.25em}
\begin{multicols}{2}

\end{multicols}


\begin{thebibliography}{99}
\bibitem{Schwingerg-2} J. Schwinger, On Quantum-Electrodynamics and the Magnetic Moment of the Electron, Phys. Rev. \textbf{73}, (1948), 416, doi:10.1103/PhysRev.73.416.

\bibitem{Petermann}
A.~Petermann,
Fourth order magnetic moment of the electron
Helv. Phys. Acta \textbf{30} (1957), 407,
doi:10.5169/seals-112823.

\bibitem{Sommerfeld}C. Sommerfield, Magnetic Dipole Moment of the Electron, Phys. Rev. \textbf{107}, (1957) 328.

\bibitem{LaportaRemiddi}
S.~Laporta and E.~Remiddi,
The Analytical value of the electron (g-2) at order $\alpha^{3}$ in QED,
Phys. Lett. B \textbf{379} (1996), 283,
doi:10.1016/0370-2693(96)00439-X.

\bibitem{Laporta}
S.~Laporta,
High-precision calculation of the 4-loop contribution to the electron g-2 in QED,
Phys. Lett. B \textbf{772} (2017), 232,
doi:10.1016/j.physletb.2017.06.056.

\bibitem{Kinoshita}
T. Aoyama, T. Kinoshita and M. Nio,
Revised and improved value of the QED tenth-order electron anomalous magnetic moment,
Phys. Rev. D \textbf{97} (2018), 036001,
doi:10.1103/PhysRevD.97.036001.

\bibitem{Volkov}
S.~Volkov,
Calculating the five-loop QED contribution to the electron anomalous magnetic moment: Graphs without lepton loops,
Phys. Rev. D \textbf{100} (2019) no.9, 096004
doi:10.1103/PhysRevD.100.096004.

\bibitem{Pred}
P.~Cvitanovic,
Asymptotic Estimates and Gauge Invariance,
Nucl. Phys. B \textbf{127} (1977), 176
doi:10.1016/0550-3213(77)90357-1.

\bibitem{BDS}
Z. Bern, D. C. Dunbar and T. Shimada,
String-Based Methods in Perturbative Gravity,
Phys. Lett. B \textbf{312} (1993), 277,
doi:10.1016/0370-2693(93)91081-W.

\bibitem{BK1}
Z. Bern and D. A. Kosower, Color decomposition of one-loop amplitudes in gauge theories, Nucl. Phys. B, {\bf 362}, (1991), 389, doi:10.1016/0550-3213(91)90567-H.

\bibitem{BK2}
Z. Bern and D. A. Kosower, The computation of loop amplitudes in gauge theories, Nucl. Phys. B {\bf 379},(1992), 451, doi:10.1016/0550-3213(92)90134-W.

\bibitem{Strass1}
M. J. Strassler, Field theory without Feynman diagrams: One loop effective actions, Nucl. Phys. B {\bf ,385},(1992), 145, doi:10.1016/0550-3213(92)90098-V.

\bibitem{Feyn1}
R. P. Feynman, Mathematical Formulation of the Quantum Theory of Electromagnetic Interaction, Phys. Rev. {\bf 80}, (1950), 440, link.aps.org/doi/10.1103/PhysRev.80.440.

\bibitem{Feyn2}
R.P. Feynman, An Operator Calculus Having Applications in Quantum Electrodynamics, Phys. Rev. {\bf 84}, (1951), 108, link.aps.org/doi/10.1103/PhysRev.84.108. 

\bibitem{KLT}
H. Kawai, D. C. Lewellen and S. H. H. Tye, A Relation Between Tree Amplitudes of Closed and Open Strings, Nucl. Phys. \textbf{B269} (1986) 1-23, doi:10.1016/0550-3213(86)90362-7.

\bibitem{BernKLT}
Z.~Bern,
Perturbative quantum gravity and its relation to gauge theory,
Living Rev. Rel. \textbf{5} (2002), 5,
doi:10.12942/lrr-2002-5.

\bibitem{BCF}
R.~Britto, F.~Cachazo and B.~Feng,
New recursion relations for tree amplitudes of gluons,
Nucl. Phys. B \textbf{715} (2005), 499,
doi:10.1016/j.nuclphysb.2005.02.030.

\bibitem{BCFW}
R.~Britto, F.~Cachazo, B.~Feng and E.~Witten,
Direct proof of tree-level recursion relation in Yang-Mills theory,
Phys. Rev. Lett. \textbf{94} (2005), 181602,
doi:10.1103/PhysRevLett.94.181602.

\bibitem{BDDK}
Z.~Bern, L.~J.~Dixon, D.~C.~Dunbar and D.~A.~Kosower,
One loop n-point gauge theory amplitudes, unitarity and collinear limits,
Nucl. Phys. B \textbf{425} (1994), 217,
doi:10.1016/0550-3213(94)90179-1,
[arXiv:hep-ph/9403226 [hep-ph].

\bibitem{CHY}
F.~Cachazo, S.~He and E.~Y.~Yuan,
Scattering equations and Kawai-Lewellen-Tye orthogonality,
Phys. Rev. D \textbf{90} (2014) no.6, 065001,
doi:10.1103/PhysRevD.90.065001.

\bibitem{BCJ}
Z.~Bern, J.~J.~M.~Carrasco and H.~Johansson,
New Relations for Gauge-Theory Amplitudes,
Phys. Rev. D \textbf{78} (2008), 085011,
doi:10.1103/PhysRevD.78.085011.

\bibitem{Scherk}
J.~Scherk,
Zero-slope limit of the dual resonance model,
Nucl. Phys. B\textbf{31} (1971), 222,
doi:10.1016/0550-3213(71)90227-6.

\bibitem{Yoneya}
T.~Yoneya,
Quantum gravity and the zero slope limit of the generalized Virasoro model,
Lett. Nuovo Cim. \textbf{8} (1973), 951,
doi:doi:10.1007/BF02727806.

\bibitem{Schwarz}
J.~Scherk and J.~H.~Schwarz,
Dual Models for Nonhadrons,
Nucl. Phys. B \textbf{81} (1974), 118,
doi:10.1016/0550-3213(74)90010-8.


\bibitem{ChrisRev}
C.~Schubert,
Perturbative quantum field theory in the string inspired formalism,
Phys. Rept. \textbf{355} (2001), 73,
doi:10.1016/S0370-1573(01)00013-8.

\bibitem{Mansfield}
P.~Mansfield,
String theory,
Rept. Prog. Phys. \textbf{53} (1990), 1183,
doi:10.1088/0034-4885/53/9/002.

\bibitem{UsRep}
J.~P.~Edwards and C.~Schubert,
Quantum mechanical path integrals in the first quantised approach to quantum field theory,
[arXiv:1912.10004 [hep-th]].

\bibitem{103}
O.~Corradini, C.~Schubert, J.~P.~Edwards and N.~Ahmadiniaz,
Spinning Particles in Quantum Mechanics and Quantum Field Theory, 
[arXiv:1512.08694 [hep-th]].

\bibitem{UsSigma}
J.~P.~Edwards, C.~M.~Mata, U.~M\"uller and C.~Schubert,
New Techniques for Worldline Integration,
SIGMA \textbf{17} (2021), 065,
doi:10.3842/SIGMA.2021.065,
[arXiv:2106.12071 [hep-th]].

\bibitem{BGK}
F.A. Berends, W.T. Giele and H. Kuijf,
On relations between multi-gluon and multi-graviton scattering,
Phys. Lett. \textbf{B211}, (1988), 91, 
doi:10.1016/0370-2693(88)90813-1.

\bibitem{GSB}
M.~B.~Green, J.~H.~Schwarz and L.~Brink,
N=4 Yang-Mills and N=8 Supergravity as Limits of String Theories,
Nucl. Phys. B \textbf{198} (1982), 474,
doi:10.1016/0550-3213(82)90336-4.

\bibitem{DunNor}
D.~C.~Dunbar and P.~S.~Norridge,
Calculation of graviton scattering amplitudes using string based methods,
Nucl. Phys. B \textbf{433} (1995), 181,
doi:10.1016/0550-3213(94)00385-R.

\bibitem{Ahmadiniaz:2021ayd}
N.~Ahmadiniaz, F.~M.~Balli, O.~Corradini, C.~Lopez-Arcos, A.~Q.~Velez and C.~Schubert,
Manifest colour-kinematics duality and double-copy in the string-based formalism,
[arXiv:2110.04853 [hep-th]].

\bibitem{ElvangHuang}
H.~Elvang and Y.~t.~Huang,
Scattering Amplitudes,
[arXiv:1308.1697 [hep-th]].

\bibitem{WLGrav1}
F.~Bastianelli and A.~Zirotti,
Worldline formalism in a gravitational background,
Nucl. Phys. B \textbf{642} (2002), 372,
doi:10.1016/S0550-3213(02)00683-1.

\bibitem{WLGrav2}
F.~Bastianelli and R.~Bonezzi,
One-loop quantum gravity from a worldline viewpoint,
JHEP \textbf{07} (2013), 016,
doi:10.1007/JHEP07(2013)016.

\end{thebibliography}
\end{document}